\documentstyle[12pt,epsf,epsfig]{article}
\def\1/2{\raise.5mm\hbox{{${\scriptstyle{1\over 2}}$}}}
\def\3/2{\raise.5mm\hbox{{${\scriptstyle{3\over 2}}$}}}

\def\5/2{\raise.3mm\hbox{{${\scriptstyle{5\over 2}}$}}}

\def\7/2{\raise.5mm\hbox{{${\scriptstyle{7\over 2}}$}}}
\def\s/2{\raise.5mm\hbox{{${\scriptstyle{s\over 2}}$}}}

\def\s{\smallskip}

\def\b{\bigskip}
\def\bb{\bigskip\bigskip}

\def\sqr#1#2{{\vcenter{\vbox{\hrule height.#2pt
 \hbox{\vrule width.#2pt height#1pt \kern#1pt
 \vrule width.#2pt} \hrule height.#2pt}}}}

\def\operp{\hbox{${\kern+.25em{\bigcirc}
\kern-.85em\bot\kern+.85em\kern-.25em}$}}

\def\lsim{\;\raise0.3ex\hbox{$<$\kern-0.75em\raise-1.1ex\hbox{$\sim$}}\;}
\def\gsim{\;\raise0.3ex\hbox{$>$\kern-0.75em\raise-1.1ex\hbox{$\sim$}}\;}
\def\no{\noindent}

\def\ve{\vfill\eject}
\def\rdots{\mathinner{\mkern1mu\raise1pt\vbox{\kern7pt\hbox{.}}\mkern2mu
 \raise4pt\hbox{.}\mkern2mu\raise7pt\hbox{.}\mkern1mu}}

\def\e e{$e^+ e^-$ }

\def\to{\rightarrow}

\begin{document}
\pagestyle{empty}
\begin{flushright}
CERN-TH/98-49\\
UCLA/98/TEP/1\\
hep-th/9802126
\end{flushright}
\vspace*{5mm}
\begin{center}
{\bf GAUGE FIELDS AS COMPOSITE BOUNDARY EXCITATIONS}\\
\vspace*{1cm}
{\bf Sergio Ferrara}\\
Theoretical Physics Division, CERN\\
CH-1211 Geneva, Switzerland\\
and\\
{\bf Christian Fr\o nsdal}\\
Physics Department, University of California\\
Los Angeles, CA 90090-1547, U.S.A.\\
\vspace*{2cm}
{\bf ABSTRACT}\\
\end{center}
\vspace*{5mm}
We investigate representations of the
conformal group that describe ``massless" particles in the interior and at the
boundary of anti-de Sitter space.

It turns out that massless gauge excitations in anti-de Sitter are gauge
``current" operators at the boundary.  Conversely, massless excitations
at the boundary are topological singletons in the interior.  These
representations lie at the threshold of two ``unitary bounds" that apply to
any conformally invariant field theory.
Gravity and Yang-Mills gauge symmetry in anti-De Sitter is translated to
global translational symmetry and continuous $R$-symmetry of the boundary 
superconformal field theory.
\vspace*{5cm}
\begin{flushleft}
CERN-TH/98-49\\
UCLA/98/TEP/1\\
February 1998
\end{flushleft}
\vfill\eject
\setcounter{page}{1}
\pagestyle{plain}

\section*{Introduction}

In recent times, new evidence [1,2,3,4] has emerged for a possible
connection between
brane dynamics in $M$ and string theories and certain supergravity theories
in anti-de Sitter geometries [5].  More specifically, it has been conjectured
that a new ``duality" [1] may occur between the $p$-brane world volume theory
and anti-de Sitter supergravity in $Ad_{p+2}$, with the same underlying
superalgebra [6,7,8].

If this duality is to work, not only the superalgebra must be the same
but also ``representations" on states and on fields should be dual to each
other.

In this paper we argue that, due to the particular nature of the
$SO(p+1,2)$ conformal group, it is   possible to make such correspondence
in a meaningful group-theoretical way.

This fact was anticipated [4] when saying that the ``massless" graviton, the
``massless" gravitino and the ``massless" gauge fields of anti-De Sitter space
are composite operators, belonging to the ``supercurrent multiplet" [28]
in the boundary theory.

It is the aim of this paper to further explore this conjecture by identifying
the  representations of $SO(p+1,2)$ that characterize both theories and
making the
correspondence more precise.

It turns out, to no surprise, that representation theory has the power to
chart the   physical degrees of freedom, as well as  gauge
modes,  in both theories.  A duality dictionary emerges between
``current operators" on the boundary (brane) and elementary massless
excitations in the anti-De Sitter bulk.  More importantly, not only the
excitations, but also their interactions are predicted by this correspondence,
because of the symmetries inherent to the dual theories.

The paper is organized as follows:  In Section 1 we review some basic
aspects of the particle and field representations of $SO(4,2)$ acting as the
conformal group of space time.  The same representations will appear
when $SO(4,2)$ acts on anti-De Sitter space.

Our analysis will be largely confined to the $p=3$ case
[9] although many properties
are valid for generic $p$.

In Sections 2 a catalogue of gauge fields are given together with their
gauge properties.  In Sections 3 and 4 we identify the representations of
$SO(n-1,2)$ that appear, and in Section 5
we summarize the models of compositeness of massless particles that arise
naturally in the context. Finally, the last sections
make some suggestions for future investigations.

\section{Some Properties of Conformal Fields on Minkowski Space}

In this section we recall some properties of conformal fields in Minkowski
space $M_4   = SO(4,2)/ \\IO(3,1) \oslash O(1,1)$.

For what follows we will be interested in ``current fields" namely,
representations labelled $ (E_0,j_1,j_2)$ with $j_1 = j_2 ={s\over 2}$.
We limit our discussion to bosonic currents.

These representations are associated with ``conserved
currents" of any spin for $E_0 = 2+s$.  This is the first unitarity bound;
it can be derived from the requirement of unitarity  of the theory, by
looking at the 2-point function
of irreducible $SO(4,2)$ symmetric traceless tensors $T_{\alpha_1\ldots
\alpha_N}$ satisfying supplementary conditions[10] on the six-cone [11]
$$
{y}^\mu T_{\mu \mu_1\ldots \mu_{N-1}} = \partial^\mu T_{\mu \mu _1\ldots
\mu _{N-1}} = 0.
$$
\no An explicit realization of these tensors can be given in any conformal
invariant free field theory, in terms of free massless fields on $M_4$,
carrying  representations with $j_1j_2 = 0$ and
$E_0 = 1+j$.  Massless representations are at the unitarity bound
for representations with $j_1j_2 = 0$.

In the specific case of $N=4$ Yang-Mills theory, the conformal multiplet
contains massless fields in the representation [12]
$$
D(1,0,0) \oplus D ({\3/2},{\1/2},0 ) \oplus D ({\3/2},0,{\1/2} )\oplus
D(2,1,0)\oplus D(2,0,1 ).
$$
\no The tensor currents   are   bilinears in these fields.

In 5-dimensional anti-De Sitter space  the same group $SO(4,2)$ plays the
role of space time isometry group.
The representations  with
$E_0=2+j_1+j_2$ correspond to ``massless particles".  In particular, the
graviton, the gravitino and
the vector fields correspond to
$$
D(4,1,1),~ D(\7/2,1,{\1/2}) \oplus D(\7/2,\1/2,1),~
 D(3,{\1/2},{\1/2} )
$$
\no respectively.  The appearance of these representations in the
(4-dimensional) boundary
theory, at the interacting level, is ensured by the global symmetries which
give the conserved currents with spin 2, 3/2, and 1.  These symmetries are
global, superconformal invariance, which in particular implies the existence
of 15 vector currents [13] in the adjoint representation of $SU(4)$ [4].

We see that in  five-dimensional anti-De Sitter space time these
representations give
particles and symmetries such that  consistent interactions are possible only
as anti-De Sitter $N=8$ supergravity with   $SU(4)$ Yang-Mills.

Interestingly enough, this gives a ``duality" between ``currents" on the
$p$-brane world volume and ``massless fields" in AdS$_5$ space-time
together with their geometric interactions.

Note that this would also explain why we only get massless particles up to
spin 2 in the anti-De Sitter bulk.  It has to do with the fact that it is
not possible, in an interacting theory on the world-volume, to have conserved
tensors of spin higher than two, so that $E_0 = 2+s$, with $s>2$, is
possible in free field theory only.

It is interesting to notice that the interpretation of ``massless particles"
($E_0=1+j$) in the boundary conformal theory is that of ``singletons" in
the corresponding anti-De Sitter description.  At the level of excitations
we may think of massless particles in the bulk as bound states of two
singletons, an
idea that was put forward long ago [14].  Apparently the preposed ``duality"
between brane dynamics and anti-De Sitter supergravity gives a dynamical
framework in which to study this idea of compositeness.  

Investigation of masslessness in arbitrary dimensions has been recently the
subject of a careful analysis in Ref.~[31].

It is amazing to observe that among these massless representations, a special
role is played by spin-one constituents (photons on the bulk) and spin-one
singletons (photons on the boundary).  They correspond to
$ D(3,{\1/2},{\1/2})$ and $D(2,1,0)$ respectively.
For both of them the Casimirs of $SO(4,2)$ vanish,\footnote{The values of
the three $SO(4,2)$ Casimir operators,
for the tensor representations $D(E_0,s/2,s/2)$, are given in the following
expressions [10],
$$
 C_1 = J_{ij}J^{ij} = 2s(s+2) + 2E_0(E_0-4),$$
$$
 C_2 = \epsilon_{ijklmn}J^{ij}J^{kl}J^{mn} = 0,
$$
$$
C_3 = J_{ij}J^{jk}J_{kl}J^{li} = s(s+2)[E_0(E_0-4) +3].
$$
}
giving to them a
particular role with respect to gauge symmetry, something that will be
discussed later.

\def\N{{\scriptstyle N}}

\section{Catalogue of Gauge Fields in AdS$_{\bf n}$}

  Gauge fields are of two basic types:
(1) Massless fields that carry excitations observable in the bulk.
 (2) Topological, singleton type gauge theories that describe oscillations
on the boundary.  Some
issues will be discussed in the setting of anti-De Sitter spaces of
arbitrary dimensions, but when this becomes too general for
the convenience of the exposition we shall specialize to the case of
greatest interest: space time dimensions $n = 4$ and $n = 5$.

Instead of coordinates in the strict sense we shall use the parameters
$y_0,...,y_n$ of the hyperboloid
$$
y^2 := y_0^2 -y_1^2-...-y_{n-1}^2 + y_n^2 = R^2 > 0.
$$
in a $n+1$-dimensional pseudo-euclidean space. All the fields will be taken
to be defined in $y^2 >1$, by fixing the degree of
homogeneity to the most convenenient value. This allows the boundary at
infinity to be identified with the cone
$y^2 = 0$, and the degree of homogeneity then coincides with the degree of
the field in the radial variable
$$
r = \sqrt{y_1^2 + ... + y_{n-1}^2}
$$
as $r \rightarrow \infty$; a field that is homogeneous of degree $\N$ will
behave as $r^\N$ near the boundary at infinity.
The space time symmetry group is $SO(n-1,2)$ in the bulk. On the boundary,
that can be identified with
either AdS$_{n-1}$ or $n-1$-dimensional Minkowski space, it acts as the conformal group of that space.
The angle in the plane of $(y_0,y_n)$ is identified wit
h the time, and the associated generator of rotations, with the
energy. The function
$$
y_+ := y_0 + i y_n = Ye^{it},~~ Y := \sqrt{y_0^2 + y_5^2}\eqno(2.1)
$$
carries one unit of energy.

  \b

\no{\it 2.0. Scalar and spinor gauge fields in AdS$_n$.}

The set of solutions of the scalar wave equation on AdS$_n$ normally
includes the
modes of an irreducible, highest weight representation of $SO(n-1,2)$. But
there is a very special case,
when the mass parameter is fixed to the value that makes some of the
solutions logarithmic, in which the
situation is quite different. In this case there appears a subspace of
gauge modes and the representation induced on it is
non-decomposable. The gauge modes are characterized only by a slower
fall-off at spatial infinity, and not by any local property.
For this reason there is no local, gauge invariant interaction.
Interactions consistent with unitarity are possible only
at the boundary at infinity, where the gauge modes disappear. In other
words, one has a topological gauge theory.
This phenomenon was first detected in the case of 4 space time dimensions
[15].  The general case of n-dimensional space time
was discussed recently by us, though we were primarily concerned with the
5-dimensional case [4].

There is a spinorial analogue [16]. In four dimensions the scalar and
spinor topological gauge fields were dubbed singletons,
after the peculiarly degenerate representations of $SO(3,2)$ that are
carried by the space of one-particle states. But it is the
topological nature of the gauge field that gives these theories their
appeal, and we shall use the term singleton
exclusively for this type of topological field theories.

The singletons associated with scalar and spinor fields have a very
important role to play. The scalar case has been explained
too recently for another review at this place. The spinor case has been
analyzed in detail only in 3 [17,18,19] and 4 [16]
 dimensions,  but this will not deter us from making use of them, since as
we shall see that the information that is required
is obtained easily from representation theory.

 \bb

\no{\it 2.1 Vector gauge fields on AdS$_{n}$.}

To construct  a vector field that transforms irreducibly under AdS$_n$ one
imposes all possible invariant conditions on it,
$$
(y^2\partial^2 - \kappa)A = 0,~~ y\cdot A = 0,~~\partial\cdot A = 0,\eqno(2.2)
$$ and
$$
y\cdot\partial A = \N A.\eqno(2.3)
$$
The parameters $\kappa$ and $\N$ will be fixed presently.

The essential feature that characterizes all gauge theories is the
existence of invariant, uncomplemented (precisely:
not invariantly complemented) subspaces. In the case at hand such a
subspace exists whenever the parameters
are such that the system (2.2), (2.3) admits solutions of the form
$$
A _\mu = (y^2\partial_\mu - ay_\mu)\Lambda,\eqno(2.4)
$$
for some value of the parameter $a$. One easily checks that this is the
case if and only if
$$
 \kappa = (\N + 1) (\N + n-2),~~ a = \N-1.
$$
Since the degree of homogeneity is arbitrary it will be convenient to
choose it so that $\kappa = 0$; this gives us two
possibilities,
$$
(1) ~~~\N = -1,~~~
(2)~~~\N = 2-n.
$$
\no (1) This possibility was analyzed in [4], in the case $n = 5$. It leads
to a topological gauge theory on AdS$_5$. The
fields fall off as $r^{-1}$ at infinity and on the boundary they give the
excitations of ordinary, four dimensional, conformally
invariant Maxwell theory. The theory is ``massless on the boundary".

\no (2) This is the theory  that, in a dimension higher than 4, is usually
called massless; to
be more precise we shall say that it is ``massless in the bulk". The
5-dimensional case was examined
by Gunaydin and Marcus [12].

\b
Both theories have meaningfull boundary values at infinity. As we said,
this manifold can be identified with the cone $y^2 = 0$.
In the case $n = 5$ it is the familiar Dirac cone in six dimension, the
conformal completion of Minkowski space. In this
four-dimensional setting both gauge theories were discussed in [20], where
the second type of gauge theory was called
``current type". Indeed, it is a fascinating fact that the boundary values
of this type of field, that describes massless
particles in five dimensions,   are of the type of a current, and thus
naturally associated with a composite operator, in
four dimensions.

\b
 \b
\no {\it 2.2. Tensor gauge fields in AdS$_n$.}

The procedure is the same: One requires
$$
(y^2\partial^2 - \kappa)g = 0,~~ g_{\mu\nu} = g_{\nu\mu}, ~~g_{\mu\mu} =
0,~~y\cdot g = 0,~~ \partial \cdot g = 0,
\eqno(2.5)
$$
and
$$
y\cdot \partial g = \N g.
$$
A gauge field,
$$
g_{\mu\nu} = (y^2\partial_\mu - ay_\mu)\Lambda_\nu + (\mu,\nu)\eqno(2.6)
$$
satisfies all these equations if and only if
$$
\kappa =  \N(\N + n-1), ~~ a = \N -2.
$$
Again we set $\kappa = 0$ and find two possibilities,
$$
(1)~~~ \N = 0, ~~~ (2)~~~ \N = 1-n.
$$
(1) This field theory has been examined in 4 dimensions, where it is
linearized AdS$_4$
gravity [21], hence there (in four dimensions) it is massless in the bulk.
It is part of the $OSp(8/4)$ massless graviton
supermultiplet [22].

\no (2)  The five-dimensional case was examined in [12] and interpreted in
terms of massless,
5-dimensional gravitons; it is a part of the massless $SU(2,2/4)$ graviton
supermultiplet.

Fields of other tensorial structure also admit gauge subspaces, see [21].

\section{ Modes and representations}

Physical states are associated with those unitary representations of
$so(n-1,2)$ that  have energy spectra bounded from
below; that is, highest weight representations. For every field theory we
must identify a set of solutions of the field
equations on which the natural (geometric) action of the group induces a
representation of this type.
In a true gauge theory  this representation is not induced on the field
modes themselves, but on
equivalence classes, the ignorable modes being gauge modes. The field
representation is thus
non-decomposable. We shall now identify these non-decomposable
representations.    It will be done
by examination of the modes of lowest energy (highest weight).
\b
\no{\it 3.0. Scalar and spinor fields.}

The scalar field mode (solution of the scalar field equation) of lowest
energy has the form
$$
({1\over y_+})^{E_0}.
$$
It is the highest weight vector of the representation $D(E_0,\vec 0)$ with
lowest energy $E_0$ and weight zero on $so(n-1)$.
As one applies energy raising operators to this mode one finds that, if
${n-1\over 2} - E_0 =
1\,$, then there appears a mode of the form
$$
y^2({1\over y_+})^{ E_0+2}f(\vec y).
$$
Since $y^2$ is an invariant, this mode belongs to an invariant subspace of
field modes that fall off more quickly at infinity.
Its lowest energy is $  E_0+2$ and the total representation is
$$
D(E_0,\vec 0) \to D( E_0+2,\,\vec 0).
$$

Spinor fields show a similar phenomenon. The representations are, in the
case of low dimensional space times, as follows
$$~
n = 3,~~~D(\1/2,0) \to D({\3/2},0),
$$
$$
n = 4, ~~~~D(1,\1/2) \to D(2,\1/2),
$$
$$
n = 5,~~~~D({\3/2},\1/2,0) \to D(\5/2,0,\1/2).
$$
The notation for the weights is explained below.
\b
\no{\it 3.1. Vector fields.}

The simplest highest weight   vector field mode is
$$
A\cdot z = z_+(y_+)^\N.\eqno(3.1)
$$
It is homogeneous, harmonic and divergenceless, but not transverse ($y\cdot
A \neq 0$).
The simplest transverse mode is
$$
 A\cdot z = (\vec y\,z_+ - \vec z\,y_+)(y_+)^{\N-1}.\eqno(3.2)
$$
This is the highest weight vector of the representation
$$
D(E_0, F),~~E_0 = -\N,\eqno(3.3)
$$
where the number $E_0$ is the $o(2)$ weight, the lowest energy, and F is
the highest weight of the fundamental representation
of
$so(n-1)$.
This is not always the only highest weight mode, but it will be useful to
examine it. First, we notice that it is a gauge
mode if and only if $\N = -1$: $$
 A\cdot z = (\vec y\,z_+ - \vec z\,y_+)(y_+)^{-2} = z \cdot\partial(-\vec
y/y_+).\eqno(3.4)
$$
So we must distinguish two cases.

(1) The case $\N = -1$. The  mode (3.2) is pure gauge, but there is another
highest weight mode,
$$
A\cdot z = (\vec y\wedge \vec z)(y_+)^{-2},\eqno(3.5)
$$
that is not. Actually, this is not an absolute highest weight, but is
highest weight modulo a subspace of gauge fields. Applying
the energy lowering operators to (3.5) we obtain (3.4); which, being a
gauge mode, can not give us back (3.5) on the application
of energy raising operators. So we have a representation that includes
$$
  D(2,F\wedge F) \to D(1,F).\eqno(3.6)
$$
Now we may hope that the quotient representation $D(2,F\wedge F)$ (the one
that is of interest!) is unitary; that can be
true only in very low dimension. Consider the simplest cases.

Space time dimension 3. The group is $SO(2,2)$. This case was investigated
by Gunaydin [17] and also in [18,19].  There are
several versions of three-dimensional Maxwell theory, but all of them
contain $D(2,0)$. (The fundamental representation of $o(2)$
is 2-dimensional and the representation $F\wedge F$  is the identity
representation.)
Among them, one is massless in the bulk;
others are topological and related to conformal field theories in two
dimensions [19].

Space time dimension 4. The irreducible representations of $so(3)$ are
labelled by a half-integer $j$ and the highest weight is
$j$. The representation (3.6) becomes
$$
D(2,1) \to D(1,1);\eqno(3.7)
$$
it describes the physical and gauge sectors of one version of Maxwell
theory on AdS$_4$ [23].

Space time dimension 5. The   representations of $so(4) = su(2) \otimes
su(2)$ have highest weights $j_1,j_2$, labelled by
two half-integers. The representation (3.6) becomes
$$
[D(2,1,0) \oplus D(2,0,1)]\to D(1,\1/2,\1/2),\eqno(3.8)
$$
which is the representation of $so(4,2)$ associated with electrodynamics
{\it in 4 dimensions!} In five dimensions it is a
topological singleton gauge theory [4]. On the  boundary at infinity it becomes
conformal Maxwell theory in a four-dimensional space that is the conformal
completion of both
Minkowski and anti De Sitter space.

Higher dimension. We suspect that both factors in the representation (3.6)
are non-unitary for $n > 6 $.

 (2) The case $\N = 2-n$. This case is simpler, for now the mode of
absolutely lowest energy, (3.2), is ``physical" (not a gauge
field). What may happen, however,  is that the space for which (3.3) is a
cyclic vector (for the natural action
of $so(n-1,2)$ on the field) may include gauge modes. Actually, by a law of
nature, this usually does happen whenever it can
happen. To understand when it can happen, we must turn to the theory of
highest weight representations (next section); to find out
if it  actually does happen in a particular context we must do some
calculations. In fact, it is easy to see that the gauge mode
in question is (2.4) with
$$
\Lambda = (y_+)^{\N-1}.
$$
We discuss the lowest dimensions.
\b
Space time dimension 3. Here $\N = -1$, which reduces to the case already
discussed, and which explains the proliferation of
versions of three-dimensional Maxwell theory [19].

Space time dimension 4. Now $\N = -2$. The representation generated from
(3.2) is
$$
D(2,1) \to D(3,0).\eqno(3.9)
$$
The theory is an alternative version of anti De Sitter electrodynamics
[23]. To get a theory that is conformally invariant one
must combine (3.9) with (3.7).

Space time dimension 5. Here $\N = -3$ and the mode (3.2) is the highest
weight vector of
$$
D(3,\1/2,\1/2) \to D(4,0,0).
$$
This is the representation of massless Maxwell theory in five dimensions;
that is, it is massless in
the bulk. At infinity it becomes the current type gauge theory already
mentioned in [20].

\section{ Non-decomposable representations}

To get an overview of the possibilities, and to understand the structural
relationship between
gauge theories in various dimensions, we must consult the theory of
non-decomposable,
highest weight representations of $so(n-1,2)$. It is possible to deal with
all dimensions
together, in a uniform manner, but since the picture changes radically with
$n$, especially
for the low values of $n$ that are of primary interest, we shall consider
each case in turn.
We will here consider the 3, 4 and 5-dimensional cases relevant   for
string, membrane and three-brane
horizons, respectively.

 {\it 4.1. Space time dimension 3.} The group $SO(2,2)$ is not simple, and
its representation
theory reduces to that of $SO(2,1)$ [17]. The most important representation
contains 9 subfactors,
including the identity representation [19]. The associated gauge theories
include singletons that
extend left and right moving conformal fields from the two-dimensional
boundary to the interior. The associated superconformal
algebras were discussed in [24] and [12]. We shall forego a detailed
description.

Gravity in three dimensions is locally trivial but very interesting
globally [25].
\
 {\it 4.2. Space time dimension 4.} The group $SO(3,2)$ is of rank 2, so
that an accurate picture
can be given of its weight lattice. In Fig. 1 the  coordinates are the
energy $E$ and the spin $j$.
To each point with coordinates $(E_0,j)$ ($2j$ a non-negative integer)
there is a highest weight
representation with weights in an area limited by straight lines rising at
45 degrees, and the
vertical axis. The two dotted lines are perpendicular to the roots and pass
through the point
(\3/2,-\1/2), the coordinates of one half the sum of the positive roots.
The following
irreducible, highest representations are unitary.
$$
D(E_0,j),~E_0 \geq j+1, j\geq 1;
$$
$$
D(E_0,j),~E_0 \geq j + \1/2, j = 0,\1/2.
$$

The rule that governs non-decomposable representations is very simple. In
order that one
irreducible, highest weight module extend another, two conditions must be
fullfilled. (1) First, it is necessary that their
highest weights be related to each other by an element of the group of
transformations, the Weyl group, that is
generated by the reflections through the dotted lines (the Weyl planes). It
is clear that, in most cases this implies that one
or the other lies below the limit imposed by unitarity. (2) The difference
between the two highest weights must belong to the
lattice generated by the noncompact roots. The possibilities include
$$
D(\1/2,0) \to D({\scriptstyle {5\over 2}},0),~~D(1,\1/2) \to
D(2,\1/2);\eqno(4.1)
$$
these are the singletons, topological gauge theories (scalar and spinor
fields) in four dimensions, the first topological
gauge theories to be discovered [15]. Vector gauge theories
(electrodynamics) make use of
$$
D(2,1) \to D(3,0),~~ D(2,1) \to D(1,1),\eqno(4.2)
$$
both of which have been mentioned already. Massless spinor fields are
described by $D(\3/2,\1/2)$; there is no extension and no
gauge theory. Rarita-Schwinger theory is a gauge theory of ordinary
massless particles in the bulk.

Linearized gravity in four dimensions [21] is a gauge theory that includes
tensor gauge fields. It
employs the following  non-decomposable representations
$$
[D(3,2) \oplus D(-1,1)] \to D(0,2), ~~ D(3,2) \to D(4,1).\eqno(4.3)
$$
The relative positions of the highest weights vis-\'a-vis the Weyl planes
can be observed in Fig.~1.

\no{\it 4.3. Space time dimension 5.} 
The weight lattice for $so(4,2)$ can
not be illustrated so
easily by a plane figure, but Fig. 2 is an attempt. 

\hglue2.0cm
\epsfig{figure=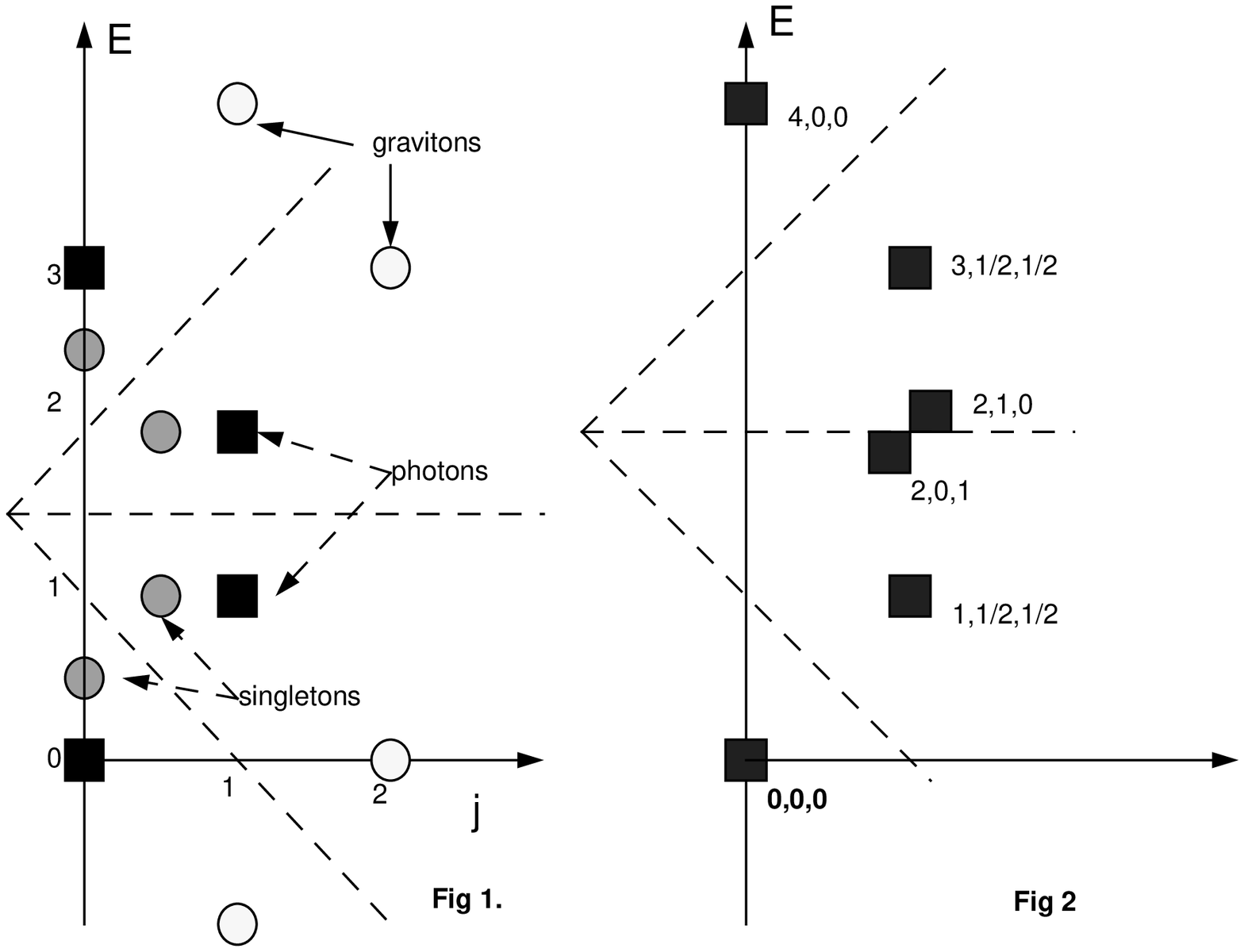,width=12cm}

The representations of
vector gauge theories
are here shown, with arrows indicating extensions. There is no general
theory of multiple
extensions, but here is an example that is worth noticing. The
representation that
is carried by the physical and the gauge sector of five-dimensional Maxwell
theory is
 fairly simple,
$$
D_1 = D(3,\1/2,\1/2) \to D(4,0,0).\eqno(4.4)
$$
Conformal electrodynamics, on the four-dimensional boundary, contains
$$
D_2 = \bigl(D(2,1,0) \oplus D(2,0,1) \oplus {\rm Id}\bigr) \to
D(1,\1/2,\1/2).\eqno(4.5)
$$
But in the extension of this latter theory, to a vector gauge theory of
singleton type on AdS$_5$,
one finds [4] the monstrous representation
$$
D_2 \to D_1
\eqno(4.6)
$$
that contains all six of the representations related among themselves by
the Weyl group. Here the
modes of five-dimensional electrodynamics appear as gauge modes of the
singleton vector
gauge theory. The non-existence of gauge invariant local interactions of
these singletons in the
bulk can thus be interpreted as a basic incompatibility between the two
types of vector gauge excitations.

An observation that we find even more fascinating is the fact that the
photons of five-
dimensional electrodynamics have boundary values at infinity that have the
characteristics
of currents rather than of massless field. Currents are composite operators
and this points to
a type of duality between composite states on the boundary and massless
fields in the bulk.

There is also the extension
$$
D(1,0,0) \to D(3,00);
$$
It is associated with a scalar singleton field in five dimensions that has
recently been examined [4]. The
spinorial analogue has not yet been studied, but it is clear that it
carries the representation
$$
D(\3/2,\1/2,0) \to D(\5/2,0,\1/2)
$$
and/or its helicity conjugate. On the boundary the gauge sector disappears
and there remains the representation
associated with massless spin-\1/2 particles in four dimensions, which is
not a gauge theory.

\section{Composite operators and particles}

Representation theory points to interesting properties of composite operators.
The  first example is the observation [14] that the singleton
representations of AdS$_4$ are related to the massless
representations of the same group by
$$
\bigl(D(\1/2,0) \oplus D(1,\1/2)\bigr) \otimes  \bigl(D(\1/2,0) \oplus
D(1,\1/2)\bigr)  = \bigoplus_s D(s+1,s),
$$
where the sum runs over all spins: $s = 0, \1/2, 1, ...$~. The direct sum
$D(\1/2,0) \oplus D(1,\1/2)$ extends to a
representation of the
AdS$_4$ supersymmetry algebra $osp(4/1)$, and the direct sum on the right
to a direct sum of massless representations of the same
algebra.

In general one has
$$
D(E_0, ...) \otimes D(E'_0,...) = D(E_0 + E'_0,...) \oplus ...~.
$$
In higher dimensions the interesting cases are, first,
\b
$$
(1)~~~~ D(2, F\wedge F)^{\otimes 2} = D(4,\vec w) \oplus ... ~.
$$
The first term is massless in the bulk if the space time dimension is 5;
in particular,
$$
D(2,1,0) \otimes D(2,0,1) = D(4,1,1) \oplus ...\, .
$$
 Since the Maxwell field on the boundary extends to a
singleton in the bulk, this gives a picture of 5-dimensional gravitons
composed of singletons.

Next,
$$
(2)~~~~D(2, F \wedge F) \otimes D(n-2,F) = D(n, F) \oplus ...~.
$$
The irreducible representations on the right hand side are not massless in
dimension $n$. The same applies to
$$
(3) ~~~~ D(n-2,F)^{\otimes 2} = D(2n-4, \vec 0) \oplus ...~.
$$
 So far we did not indicate a substructure for five-dimensional photons.
(Two four-dimen-sional massless vector
representations yield $D(4,0,0)$, which is pure gauge.) Nevertheless, there
is a way. Half integral spin
singletons tend to have lower energy than singletons with integral spins.
In particular we have
$$
(4)~~~~D(\3/2,\1/2,0)\otimes D(\3/2,0,\1/2) = D(3,\1/2,\1/2).
$$
Massless bosons in five dimensions are composites of spin-\1/2 singletons.

\section{Further outlook}

A remarkable property seems to be a general characteristic of the most
interesting gauge theories: they have a strong affinity
to the vacuum. Many are ``zero center" modules, by which we mean that all
the (super)  Casimir operators take the value zero.
The simplest way to detect that a highest weight representation has zero
center is to verify that its highest weight is related
to that of the trivial representation (zero) by an element of the Weyl
group, or by the existence of an extension that
involves the trivial representation, as in (4.5). Examples of zero center
modules are electrodynamics with its super symmetric
and superconformal extensions, and $N = 6$ supergravity. A more familiar
example is the scalar field in 2 dimensions. The field
theories associated with zero center modules (``spontaneously generated
field theories" [26]) all contain the zero mode that
is so important in 2-dimensional conformal field theory. The physical
implications of this, in dimensions higher than two, are
not yet clear.

It should be pointed out that the injunction against observing singletons
in the bulk is valid within the framework of
conventional, perturbative quantum field theory. At the price of departing
slightly from that context, to explore
alternate methods of quantization, it is possible view propagating and
interacting massless particles as 2-singleton bound
states with zero binding energy [ 27].

Finally, we notice that, since the $U(N)$ Yang-Mills theory on the boundary
can be compared, for large $N$, to the anti-De
Sitter bulk theory with radius $R^2/\alpha' = \sqrt{4\pi gN}$, the study of
the spectrum of the former should give information
about the spectrum of the other [1]. This is in strict analogy with
(matrix) M theory [29], the spectrum of which is
related to that of 11-dimensional supergravity; in  very recent papers [30,32]
some  progress was made in this direction.



%
%




\bb

{\bf Acknowledgements}

  S.F. is
supported in part by DOE
under grant DE-FG03-91ER40662, Task C, and by EEC Science Program
SCI$*$-CI92-0789 (INFN-Frascati).

\ve

\bigskip
{\bf References} 
\vskip.3cm

\begin{description}

\item{[1]} J. Maldacena, hep-th/9711200;\\
N. Itzhaki, J. Maldacena, J. Sonnenschein and S. Yankielowicz, hep-th/9802042.

\item{[2]} K. Sfetsos and K. Skenderis, hep-th/9711138;\\
H. Boonstra, B. Peeters and K. Skenderis, hep-th/9801076 {\it Phys. Lett.} {\bf
B411} (1997) 59;\\
S. Hyun, hep-th/9704005. 

\item{[3]} P. Claus, R. Kallosh and A. van Proeyen, hep-th/9711161;\\
 P. Claus, R. Kallosh, J. Kumar, P. Townsend and A. Van Proeyen,
hep-th/9801206.

\item{[4]} S. Ferrara and C. Fr\o nsdal, hep-th/ 9712239.

\item {[5]} See, e.g., A Salam and E. Sezgin, ``Supergravity in Diverse
Dimensions." (World Scientific) (1989) Vols. I and II.

\item {[6]} M.P. Blencowe and M.J. Duff, {\it Phys. Lett.} {\bf B203} (1988) 229;
{\it Nucl. Phys.} {\bf B310} (1988) 389;\\
M.J. Duff, {\it Class. Quant. Gravity} {\bf 5} (1988) 189;\\
E. Bergshoeff, M.J. Duff, C.N. Pope and E. Sezgin, {\it Phys. Lett.} {\bf B199}
(1988) 69.

\item {[7]} H. Nicolai, E. Sezgin and Y. Tanii, {\it Nucl. Phys.} {\bf B305} (1988) 483.

\item{[8]} C. W. Gibbons and P. K. Townsend, {\it Phys. Rev. Lett.} {\bf 71} (1993)
3754.

\item{[9]} M. G\"unaydin, L.J. Romans and N.P. Warner, {\it Phys. Lett.} {\bf 154B}
(1985) 268.

\item {[10]} S. Ferrara, A.F. Grillo and R. Gatto, {\it Ann. Phys.} {\bf76}
(1973) 161.

\item{[11]} G. Mack and A. Salam, {\it Am. Phys.} {\bf 53} (1969) 174.

\item {[12]} M. G\"unaydin and N. Marcus, {\it Class. Quant. Grav.} {\bf 2} (1985)
L11;\\
H.J. Kim, L.J. Romans and P. van Nieuwenhuizen, ``The Mass Spectrum of Chiral $N = 2
D = 10$ Supergravity on $S^5$, {\it Phys. Rev. } {\bf D32} (1985) 389;\\
 M. G\"unaydin and D. Minic, hep-th/9702047;\\
 M. Pernici, K. Pilch and P. van Nieuwenhuizen, {\it Nucl. Phys.} {\bf 259}
 (1985) 460;\\
 P. van Niewenhuizen, {\it Class. Quant. Grav.} {\bf 2} (1985) 1.

\item {[13]} P. Howe, K. Stelle and P. Townsend, {\it Nucl. Phys.} {\bf B192}
(1081) 332.

\item{[14]} M. Flato and C, Fr\o nsdal, {\it Lett.Math.Phys.} {\bf2} (1978) 421;
{\it Phys. Lett.} {\bf 97B} (1980) 236.

\item{[15]}   M. Flato and C. Fr\o nsdal, {\it J. Math. Phys.} {\bf 22} (1981) 1100.

\item {[16]} C. Fr\o nsdal, {\it Phys. Rev.} D {\bf 26} (1982) 1988;\\
 W. Heidenreich, {\it N.C.} {\bf 84} (1980) 220.

\item{[17]} M. G\"unaydin, Proc.  Trieste conf. {\it Supermembranes and
Physics in 2+1 dimensions}, Eds. M.J. Duff,
C.N. Pope, E. Sezgin, World Scientific, 1990, p 442;\\  M. G\"unaydin, B.E.
W. Nilsson,  G. Sierra and P.K. Townsend, {\it Phys. Lett.} {\bf B176} (1986) 45.

\item{[18]} M. Flato and C. Fr\o nsdal, {\it Lett. Math. Phys.} {\bf 20} (1990) 65.

\item{[19]} C. Fr\o nsdal, {\it Ann. Phys.} {\bf 22} (1991) 27.

\item {[20]} B. Binegar, C. Fr\o nsdal and W. Heidenreich, {\it J. Math. Phys.}
{\bf 24} (1983) 2828.

\item{[21]} C. Fr\o nsdal and W. Heidenreich, {\it J. Math. Phys.} {\bf 28} (1987) 215.

\item {[22]} B. de Wit and H. Nicolai, {\it Nucl. Phys.} {\bf B208} (1982) 323.

\item{[23]} C. Fr\o nsdal, {\it Phys. Rev.} D {\bf 12} (1975) 3819.

\item {[24]} M.G\"unaydin, B.E.W. Nilsson, G. Sierra and P. Townsend, {\it Phys
Lett.} {\bf 176B} (1986) 45.

\item{[25]} S. Deser, R. Jackiw and G. 't Hooft, {\it Ann. Phys.} {\bf 152} (1984)
220;\\
G.Horowitz and D. Welch, {\it Phys. Rev.}{\bf 71} (1993)
328;\\ 
N. Kaloper, {\it Phys. Rev.} D {\bf 48} (1993) 2598;\\
 A. Achicano and P. Townsend, {\it Phys. Lett.} {\bf B180} (1986) 89;\\
 E. Witten, {\it Nucl. Phys.} {B311} (1998) 46.

 \item{[26]} M. Flato and C. Fr\o nsdal, {\it Spontaneously generated Field
Theories, Zero-Center Modules, Colored Singletons and
the virtues of $N = 6$ Supergravity}. In ``Essays in Supersymmetry", (C.
Fr\o nsdal, Ed.) Reidel 1986.

\item{[27]} M. Flato and C. Fr\o nsdal, {\it Phys. Lett.} {\bf B172} (1986) 412; {\it
J. Geom. Phys.} {\bf 6} (1988) 294; {\it idem}
{\bf 6} (1989) 293;\\
 C. Fr\o nsdal,   proc.  Colloque Rideau, Paris, January 1995.

\item{[28]} S. Ferrara and B. Zumino, {\it Nucl. Phys.} {\bf B87} (1975) 207.
\item{[29]} T. Banks, W. Fischler, S. Shenker and L. Susskind, hep-th/9610043.

\item{[30]} G.T. Horowitz and H. Ooguri, hep-th/9802116.

\item{[31]} E. Angelopoulos and M. Laoues, ``Masslessness in n-dimensions",
Dijon (Univ. de Bourgogne, Lab. Gevrey) preprint, July 1997;\\
M. Laoues, ``Some properties of massless particles in arbitrary dimensions",
Dijon (Univ. de Bourgogne, Lab. Gevrey) preprint, Jan. 1998.
\item{[32]} S.S. Gubser, I.R. Klebanov and A.M. Polyakov, ``Gauge Theory
Correlators from Non-Critical String Theory, hep-th/9802109.
\end{description}

\end{document}